# Census Data Mining and Data Analysis using WEKA

Dr. Sudhir B. Jagtap,

Swami Vivekanand Institute of Technology and Management, Udgir

Maharashtra State, India

sudhir.jagtap7@gmail.com,

Dr. Kodge B. G.

Swami Vivekanand Institute of Technology and Management, Udgir

Maharashtra State, India

kodgebg@hotmail.com

*Abstract:*

*Data mining (also known as knowledge discovery from databases) is the process of extraction of hidden, previously unknown and potentially useful information from databases. The outcome of the extracted data can be analyzed for the future planning and development perspectives. In this paper, we have made an attempt to demonstrate how one can extract the local (district) level census, socio-economic and population related other data for knowledge discovery and their analysis using the powerful data mining tool Weka.*

I. DATA MINING

Data mining has been defined as the nontrivial extraction of implicit, previously unknown, and potentially useful information from databases/data warehouses. It uses machine learning, statistical and visualization techniques to discover and present knowledge in a form, which is easily comprehensive to humans [1].

Data mining, the extraction of hidden predictive information from large databases, is a powerful new technology with great potential to help user focus on the most important information in their data warehouses. Data mining tools predict future trends and behaviors, allowing businesses to make proactive, knowledge-driven decisions. The automated, prospective analyses offered by data mining move beyond the analyses of past events provided by retrospective tools typical of decision support systems. Data mining tools can answer business questions that traditionally were too time consuming to resolve. They scour databases for hidden patterns, finding predictive information that experts may miss because it lies outside their expectations. Data mining techniques can be implemented rapidly on existing software and hardware platforms to enhance the value of existing information resources, and can be integrated with new products and systems as they are brought on-line [2].

Data mining steps in the knowledge discovery process are as follows:

1. Data cleaning- The removal of noise and inconsistent data.
2. Data integration - The combination of multiple sources of data.
3. Data selection - The data relevant for analysis is retrieved from the database.
4. Data transformation - The consolidation and transformation of data into forms appropriate for mining.
5. Data mining - The use of intelligent methods to extract patterns from data.
6. Pattern evaluation - Identification of patterns that are interesting.

7. Knowledge presentation - Visualization and knowledge representation techniques are used to present the extracted or mined knowledge to the end user [3].

The actual data mining task is the automatic or semi-automatic analysis of large quantities of data to extract previously unknown interesting patterns such as groups of data records (cluster analysis), unusual records (anomaly detection) and dependencies (association rule mining). This usually involves using database techniques such as spatial indices. These patterns can then be seen as a kind of summary of the input data, and may be used in further analysis or, for example, in machine learning and predictive analytics. For example, the data mining step might identify multiple groups in the data, which can then be used to obtain more accurate prediction results by a decision support system. Neither the data collection, data preparation, nor result interpretation and reporting are part of the data mining step, but do belong to the overall KDD process as additional steps [7][8].

II. WEKA:

Weka (Waikato Environment for Knowledge Analysis) is a popular suite of machine learning software written in Java, developed at the University of Waikato, New Zealand. Weka is free software available under the GNU General Public License. The Weka workbench contains a collection of visualization tools and algorithms for data analysis and predictive modeling, together with graphical user interfaces for easy access to this functionality [4].

Weka is a collection of machine learning algorithms for solving real-world data mining problems. It is written in Java and runs on almost any platform. The algorithms can either be applied directly to a dataset or called from your own Java code [5].

The original non-Java version of Weka was a TCL/TK front-end to (mostly third-party) modeling algorithms implemented in other programming languages, plus data preprocessing utilities in C, and a Makefile-based system for running machine learning experiments. This original version was primarily designed as a tool for analyzing data from agricultural domains, but the more recent fully Java-based version (Weka 3), for which development started in 1997, is now used in many different application areas, in particular for educational purposes and research. Advantages of Weka include:

I. Free availability under the GNU General Public License
II. Portability, since it is fully implemented in the Java programming language and thus runs on almost any modern computing platform
III. A comprehensive collection of data preprocessing and modeling techniques
IV. Ease of use due to its graphical user interfaces

Weka supports several standard data mining tasks, more specifically, data preprocessing, clustering, classification, regression, visualization, and feature selection [10]. All of Weka's techniques are predicated on the assumption that the data is available as a single flat file or relation, where each data point is described by a fixed number of attributes (normally, numeric or nominal attributes, but some other attribute types are also supported). Weka provides access to SQL databases using Java Database Connectivity and can process the result returned by a database query. It is not capable of multi-relational data mining, but there is separate software for converting a collection of linked database tables into a single table that is suitable for processing using Weka. Another important area that is currently not covered by the algorithms included in the Weka distribution is sequence modeling [4].

III. DATA PROCESSING, METHODOLOGY AND RESULTS

The primary available data such as census (2001), socio-economic data, and few basic information of Latur district are collected from National Informatics Centre (NIC), Latur, which is mainly required to design and develop the database for Latur district of Maharashtra state of India.

The database is designed in MS-Access 2003 database management system to store the collected data. The data is formed according to the required format and structures.

Further, the data is converted to ARFF (Attribute Relation File Format) format to process in WEKA. An ARFF file is an ASCII text file that describes a list of instances sharing a set of attributes. ARFF files were developed by the Machine Learning Project at the Department of Computer Science of The University of Waikato for use with the Weka machine learning software. This document descibes the version of ARFF used with Weka versions 3.2 to 3.3; this is an extension of the ARFF format as described in the data mining book written by Ian H. Witten and Eibe Frank [6][9].

After processing the ARFF file in WEKA the list of all attributes, statistics and other parameters can be utilized as shown in Figure 1.

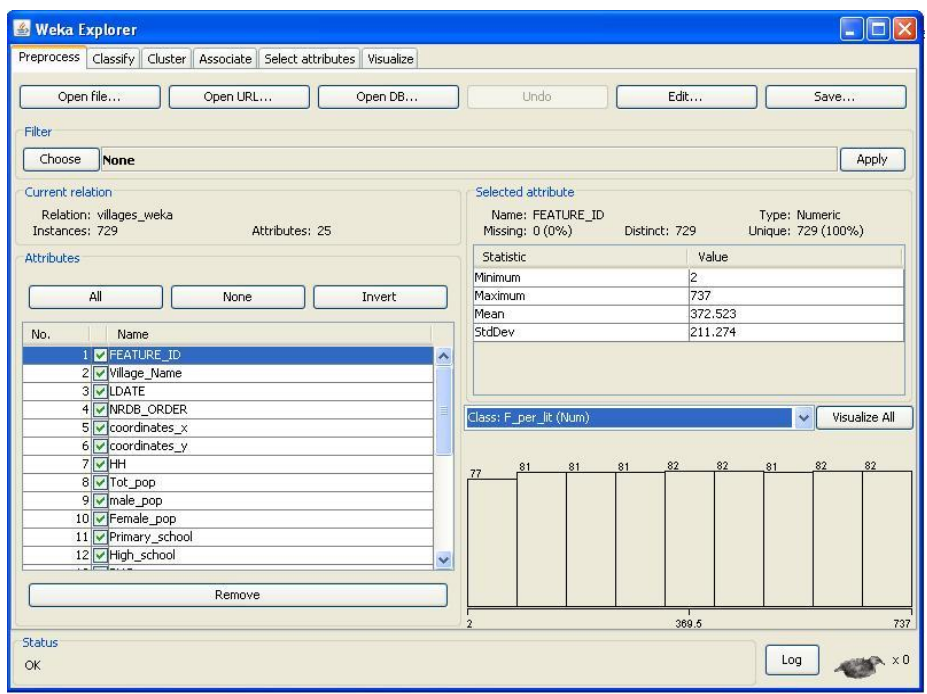

Fig.1 Processed ARFF file in WEKA.

In the above shown file, there are 729 villages data is processed with different attributes (25) like population, health, literacy, village locations etc. Among all these, few of them are preprocessed attributes generated by census data like, percent_male_literacy, total_percent_literacy, total_percent_illiteracy, sex_ratio etc.

The processed data in Weka can be analyzed using different data mining techniques like, Classification, Clustering, Association rule mining, Visualization etc. algorithms. The Figure 2 shows the few processed attributes which are visualized into a 2 dimensional graphical representation.

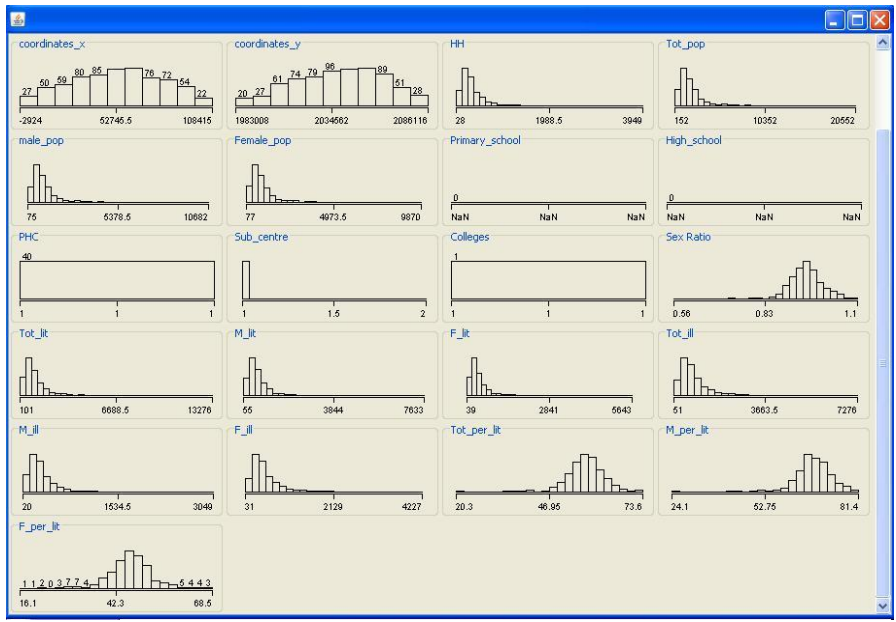

Fig. 2 Graphical visualization of processed attributes.

The information can be extracted with respect to two or more associative relation of data set. In this process, we have made an attempt to visualize the impact of male and female literacy on the gender inequality. The literacy related and population data is processed and computed the percent wise male and female literacy. Accordingly we have computed the sex ratio attribute from the given male and female population data. The new attributes like, male_percent_literacy, female_percent_literacy and sex_ratio are compared each other to extract the impact of literacy on gender inequality. The Figure 3 and Figure 4 are the extracted results of sex ratio values with male and female literacy.

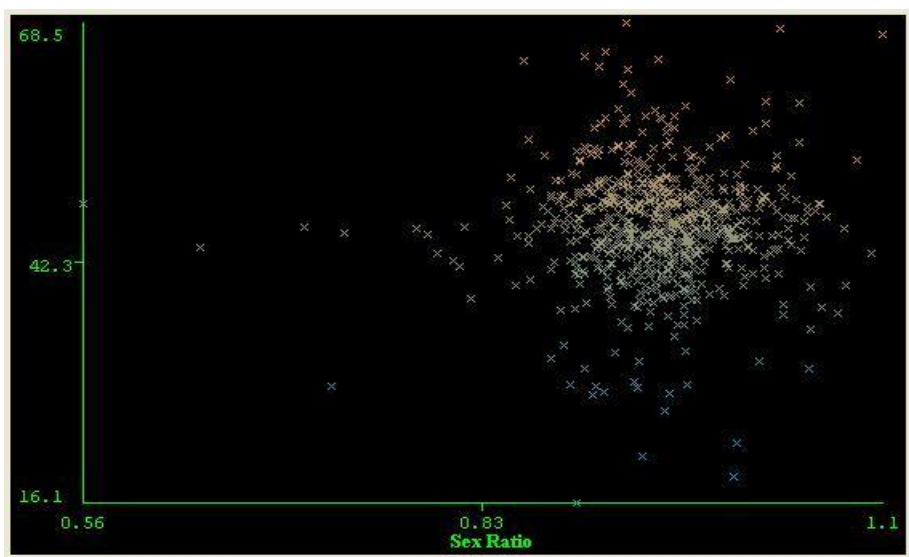

Fig. 3 Female literacy and Sex ratio values.

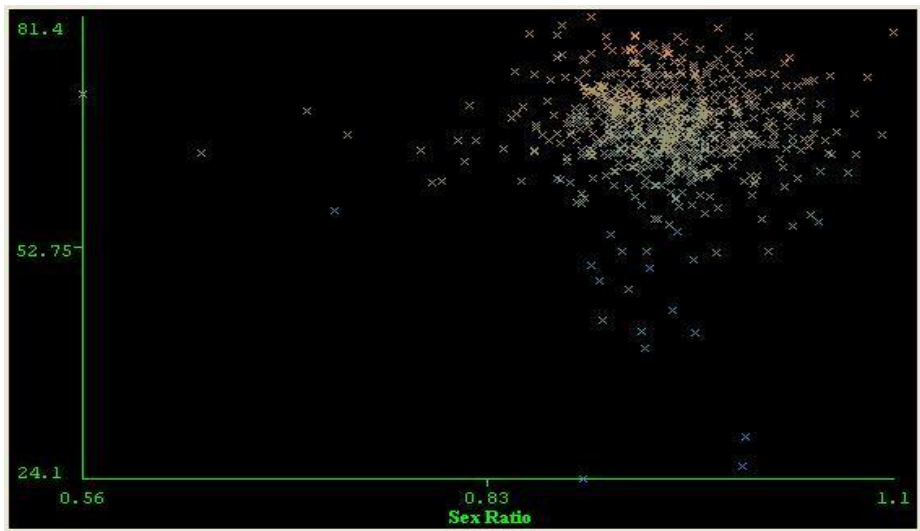

Fig. 4 Male literacy and Sex ratio values.

On the Y-axis, the female percent literacy values are shown in Figure 3, and the male percent literacy values are shown in Figure 4. By considering both the results, the female percent literacy is poor than the male percent literacy in the district. The sex ratio values are higher in male percent literacy than the female percent literacy.

The results are purely showing that the literacy is very much important to manage the gender inequality of any region.

ACKNOWLEDGEMENT:

Authors are grateful to the department of NIC, Latur for providing all the basic data and WEKA for providing such a strong tool to extract and analyze knowledge from database.

CONCLUSION

Knowledge extraction from database is become one of the key process of each and every organization for their development issues. This is not only important in the commercial industries but also plays a vital role in e-governance for future planning and development issues. This paper shows one of the small importances of Weka to utilization and analysis for census data mining issues and knowledge discovery. Its for E-Governance.

REFERENCES:

[1] Jiawei Han and Micheline Kamber, Data Mining Concepts and Techniques, 2nd ed., Morgan Kaufmann publishers, SanFrancisco, 2006.
[2] George M. Marakas, Modern Data Warehousing, Mining, and Visualization, Pearson Education, New Delhi, 2005.
[3] Michael J.A. Berry and Gordon S. Linoff, Data Mining Techniques, 2nd ed., Wiley Publishing Inc., USA, 2004.
[4] http://en.wikipedia.org/wiki/Weka_(machine_learning)
[5] http://www.cs.waikato.ac.nz/ml/weka/
[6] http://www.cs.waikato.ac.nz/ml/weka/arff.html
[7] http://en.wikipedia.org/wiki/Data_mining

[8] Margaret H. Dunham, Data Mining Introductory and Advanced Topics, Pearson Education, New Delhi, 2009

[9] http://www.iasri.res.in/ebook/win_school_aa/notes/WEKA.pdf

[10]     Sunita B Aher, Lobo LMRJ, Data Mining in Educational System using Weka, International Conference on Emerging Technology Trends (ICETT), Proceedings published by International Journal of Computer Applications® (IJCA) Number 3, 2011, pp-20-25.

[11]     http://www.cs.ccsu.edu/~markov/weka-tutorial.pdf